\documentclass{article}

\usepackage{microtype}
\usepackage{graphicx}
\usepackage{subfigure}
\usepackage{booktabs} 

\usepackage{hyperref}

\usepackage[accepted]{icml2020}

\usepackage{amsmath}

\usepackage{amsfonts}

\DeclareMathOperator*{\argmin}{argmin}
\usepackage[dvipsnames]{xcolor}
\usepackage{pifont}
\usepackage{multirow}

\newcommand{\D}{\mathbf{D}}
\newcommand{\X}{\mathbf{X}}

\newcommand{\A}{\mathbf{A}}

\newcommand{\perm}{\mathbf{P}}

\newcommand{\G}{\mathcal{G}}

\newcommand{\V}{\mathcal{V}}
\newcommand{\E}{\mathcal{E}}

\renewcommand{\algorithmicrequire}{\textbf{Input:}}
\renewcommand{\algorithmicensure}{\textbf{Output:}}

\icmltitlerunning{Learning Heat Diffusion for Network Alignment}

\begin{document}

\twocolumn[
\icmltitle{Learning Heat Diffusion for Network Alignment}




\begin{icmlauthorlist}
\icmlauthor{Sisi Qu}{kaust}
\icmlauthor{Mengmeng Xu}{ivul}
\icmlauthor{Bernard Ghanem}{ivul}
\icmlauthor{Jesper Tegner}{kaust}
\end{icmlauthorlist}

\icmlaffiliation{kaust}{Living Systems Laboratory, King Abdullah University of Science and Technology (KAUST), Saudi Arabia.}
\icmlaffiliation{ivul}{Visual Computing Center, KAUST, Saudi Arabia}

\icmlcorrespondingauthor{Jesper Tegner}{jesper.tegner@kaust.edu.sa}

\icmlkeywords{Machine Learning, ICML}

\vskip 0.3in
]



\printAffiliationsAndNotice{} 

\begin{abstract}
Networks are abundant in the life sciences. Outstanding challenges include how to characterize similarities between networks, and in extension how to integrate information across networks. Yet, network alignment remains a core algorithmic problem. Here, we present a novel learning algorithm called evolutionary heat diffusion-based network alignment (EDNA) to address this challenge. EDNA uses the diffusion signal as a proxy for computing node similarities between networks. Comparing EDNA with state-of-the-art algorithms on a popular protein-protein interaction network dataset, using four different evaluation metrics, we achieve (i) the most accurate alignments, (ii) increased robustness against noise, and (iii) superior scaling capacity. The EDNA algorithm is versatile in that other available network alignments/embeddings can be used as an initial baseline alignment, and then EDNA works as a wrapper around them by running the evolutionary diffusion on top of them. In conclusion, EDNA outperforms state-of-the-art methods for network alignment, thus setting the stage for large-scale comparison and integration of networks. 

\end{abstract}

\section{Introduction}

Considerable biological data related to signaling, metabolic, and regulatory relationships is usually represented as networks/graphs and is being generated at an increasing pace. However, such graphs tend to be challenging to process as they do not conform to grid-like structure or Euclidean geometry. 
A core algorithmic challenge in this domain is network alignment (NA), \textit{i.e.} the process in which similarities between networks (leading to node correspondences) are quantitatively identified for the purpose of merging, information transfer. The task of NA has been extensively studied over the last two decades. Protein-protein interaction (PPI) networks, where nodes are proteins that are connected with edges representing interactions, are of particular importance in biology. PPI networks contain valuable information for understanding protein functions, modular organization of cells, and system-level cellular processes \cite{hashemifar2014hubalign}. Recently, the size of PPI networks has rapidly increased  and numerous large-scale PPI maps have been produced by high throughput techniques in \citep{huttlin2015bioplex}, \citep{huttlin2017architecture}, \citep{luck2019reference}, \textit{etc}. Comparative analysis of PPI networks has therefore become a key challenge as well as serving as a vehicle for data-driven integration of PPI networks. 

\begin{figure}[t]
    \centering
    \includegraphics[trim={8cm 3cm 8cm 2.5cm},width=9cm,clip]{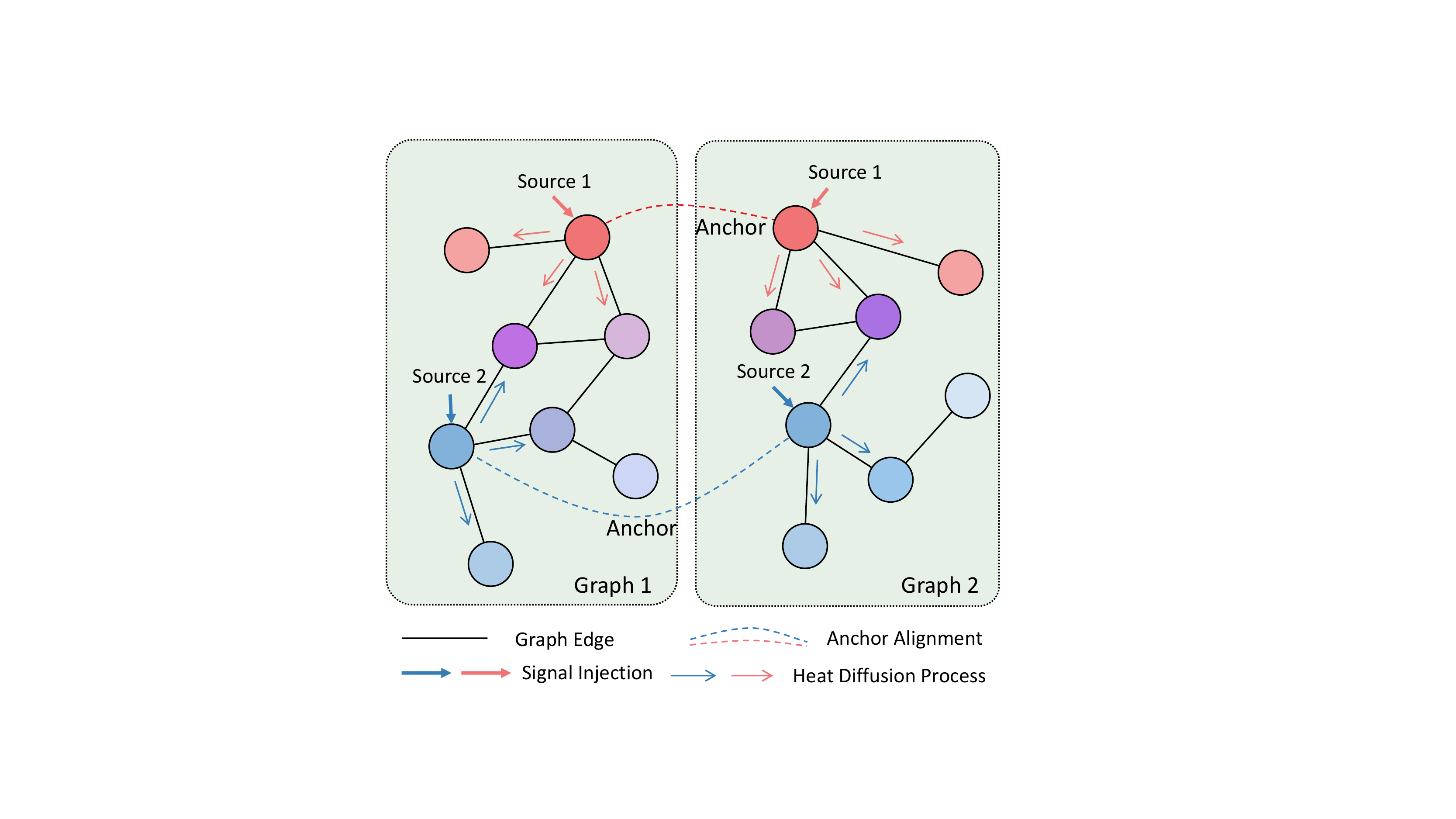}
    \small
    \vspace{-20pt}
    \caption{\textbf{Heat diffusion process from anchor node-pairs in two networks (graphs).} Specifically, in a PPI network, a node is a protein and an edge denotes an interaction. The signals of anchor nodes diffuse to all nodes, thus encoding both local structure and global structure of two networks.} 
     \vspace{-12pt}
    \label{fig:intro}
\end{figure}

\begin{figure*}[!th]
 \vspace{-5pt}
    \centering
    \includegraphics[trim={1cm 2.6cm 1cm 2.5cm},width=16cm,clip]{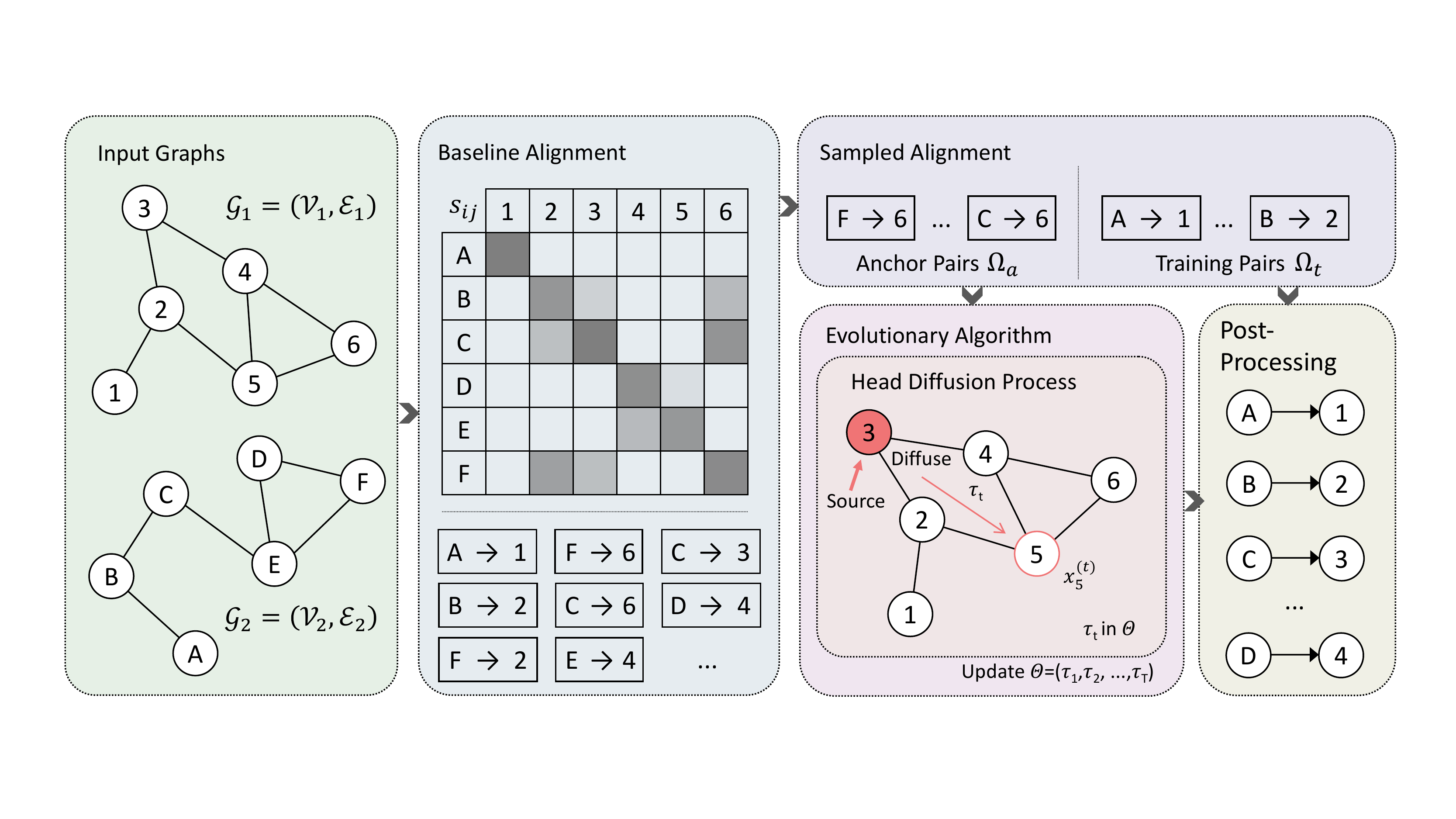}
    \small
    \vspace{-12pt}
    \caption{\textbf{Improved network alignment from evolutionary heat diffusion framework.} Using two input networks (graphs), we construct the baseline alignment based on embeddings or available alignments. Then we sample the alignment with highest precision as our anchor alignments. After applying heat diffusion using the signals of anchor alignment nodes to the whole graph $T$ times, the evolutionary algorithm can be harnessed to fine-tune the diffusion parameter $\Theta$ to further improve performance.} 
    \vspace{-8pt}
    \label{fig:framework}
\end{figure*}
Our work focuses on global network alignment (GNA) aiming at maximizing matches between entire networks first proposed by \citep{singh2008global}, with a specific application on PPI networks. GNA can be used to study interactomes and for understanding cross-species variations among PPI networks. Moreover, the GNA can facilitate the detection of functional ortohologs. Here, we propose \textbf{evolutionary heat diffusion-based network alignment (EDNA)} to address the PPI network alignment problem. There are two stages in the EDNA pipeline. First, we obtain the baseline alignment calculated from extracted PPI network representations or other network alignment techniques, and on this basis we select nodes with highly confident alignments as anchor node-pairs. Secondly, evolutionary diffusion is deployed using those anchors with injected signals to diffuse information throughout the whole graph, and then a similarity matrix between the two networks is computed to enable network matching. Specifically, the heat diffusion process is applied on source anchor nodes as illustrated in Fig. \ref{fig:intro}. Importantly, EDNA is the first flexible framework that can be employed both on representations extracted from PPI networks by a variety of embedding techniques to obtain alignments, and also on any baseline alignment directly sampled from currently available network alignment algorithms to improve its performance (Fig. \ref{fig:framework}). We demonstrate the improved performance on a PPI network dataset using four different evaluation metrics with superior robustness, scalability, and computational efficiency.




\section{Method}

The EDNA framework is illustrated in Fig. \ref{fig:framework}. 
Let $\G_1=(\V_1, \E_1)$ and $\G_2=(\V_2, \E_2)$ be two unweighted and undirected networks with node sets $\mathcal{V}_1$ and $\mathcal{V}_2$; edge sets $\mathcal{E}_1$ and $\mathcal{E}_2$, respectively. 
We construct the baseline alignment and sample the top $\gamma$ confident alignment pairs to build the training set $\Omega_t$ and anchor set $\Omega_a$ (Sec.~\ref{sec:step1}). 
We inject identical signals into paired nodes in $\Omega_a$, and diffuse it for $T$ times on $\G_1$ and $\G_2$, while the parameter $\Theta$ is optimized by evolutionary algorithms (Sec.~\ref{sec:step2}).
With diffused signal $\boldsymbol{x}_i^{(T)}$ on each node $\nu_i$, we update the baseline alignment based on the signal similarity in post-processing (Sec.~\ref{sec:post}). Finally we analyse the computation complexity in Sec.~\ref{sec:comp}.

    
\subsection{Baseline Alignment Sampling}
\label{sec:step1}
Many works have studied NA by exploring the feature representation from prior knowledge.
Each node $\nu_i\in \V$ is represented by a embedding vector $\boldsymbol{e}_i$. One could also incorporate biological features into the PPI representations (\textit{e.g.}, sequence similarity scores computed by BLAST) to enrich the encoded information.
To align the networks, we take the node representation for two nodes $\nu_i\in \V_1,\nu_j\in \V_2$, apply an exponential kernel on their Euclidean distance in the embedding space, 
and take the output as the alignment confidence score, shown in Eq.~\ref{eq:sim}.
\begin{align}\label{eq:sim}
s_{ij}=\exp(-\|\boldsymbol{e}_i-\boldsymbol{e}_j\|_2^2)
\end{align}
By doing so, we have a list of node pairs $(i,j)$ associated with confidence score $s_{ij}$ as our baseline alignment. Worth-mentioning, EDNA can be flexibly integrated with other non-embedding NA methods. Alignments from those NA methods can be directly fed into our pipeline as input.

From the baseline alignment, we respectively sample $\alpha$ and $\beta$ node pairs to form an anchor set $\Omega_a$ and a training set $\Omega_t$.
To include randomness in the alignment samples, we compute the mean $\mu_s$ and standard derivation $\sigma_s$ of the all the confidence scores $s_.$, and randomly sample $\gamma=\alpha+\beta$ confident $(i,j)$-pairs where $s_{i,j}\geq \mu_s+\sigma_s$. 

\subsection{Evolutionary Heat Diffusion Process}
\label{sec:step2}
After injecting identical signals into those paired nodes in $\Omega_a$ and simulate $T$ heat diffusion steps on both $\G_1$ and $\G_2$. The diffused value $\boldsymbol{x}_i^{(T)}$ depends on the neighbourhood of $\nu_i$, reflecting the topology information. We leverage an evolutionary algorithm to learn the diffusion parameters. 

We respectively initialize the anchor and non-anchor node values by identical signals and zeroes. 
In practice, we generate $\alpha$ unique vectors $\boldsymbol{f}_k \in \mathbb{R}^C, k=1,2,...,\alpha$ and assign them to each of the paired nodes. We keep the signals on other nodes as zero-vectors. Eq.~\ref{eq:init1} and Eq.~\ref{eq:init2} show the details to assign the initial graph signals, where $\Omega_a(k)$ designates the $k$-th node-pair in  $\Omega_a$.
\begin{align}
\forall \nu_i \in \G_1, \boldsymbol{x}_i^{(0)}=\left\{
\begin{aligned}
\boldsymbol{f}_k, ~~&\exists \Omega_a(k)=(i,j) \\
\boldsymbol{0}, ~~& \text{otherwise}
\end{aligned}
\right. 
\label{eq:init1}
\\
\forall \nu_j \in \G_2, \boldsymbol{x}_j^{(0)}=\left\{
\begin{aligned}
\boldsymbol{f}_k, ~~&\exists \Omega_a(k)=(i,j) \\
\boldsymbol{0}, ~~& \text{otherwise}
\end{aligned}
\right.
\label{eq:init2}
\end{align}
Thus, when we assign the identical signals to the $2 \alpha$ nodes, only the paired nodes have the same initial heat values. 

Then, we simulate the heat diffusion process on the corresponding graphs. In the process, the neighbouring node information is aggregated. 
The diffusion process can be formulated under different assumptions. In this paper we use Eq.~\ref{eq:diffuse} \cite{7990049}, where $\A\in \mathbb{R}^{N\times N}$ and $\D\in \mathbb{R}^{N\times N}$ are respectively the adjacency matrix and degree matrix of the graph, which has $N$ nodes. $\X^{(t)} \in \mathbb{R}^{N\times C}$ is the stacking of all the node signal $\boldsymbol{x}^{(t)}_i$ on the graph. We use the element-wise activation function $\sigma(x)=[ 1 + \exp(-x)]^{-1}$  to create non-linearity and map the value to the range $(0,1)$. $\tau_t$ is the diffusion duration in $t$-th iteration.
\begin{align}
\X^{(t+1)}=\X^{(t)}+\sigma(\tau_t (\D-\A) \X^{(t)})
\label{eq:diffuse}
\end{align}


We leverage a genetic algorithm to learn the diffusion duration $\Theta=[\tau_1,...,\tau_T]$. We take as the fitness function the node accuracy on train alignment, $\Omega_t$. 
The node accuracy $f(\Theta)$ is the portion of positive alignments (\textit{e.g.} $l==j$) over all training alignment pairs. Specifically, for each node pair $(i,j)$ in $\Omega_t$, the positive case is when the closest node $\nu_l$ in $\G_2$ of $\nu_i$ in $\G_1$ is exactly $\nu_j$, shown in Eq.~\ref{eq:fit}, where $\textbf{1}\{\}$ is the indicator function.
\begin{align}
f(\Theta)=\frac{1}{\beta}\sum_{(i,j) \in \Omega_t} \textbf{1}\{j=\argmin_l(\|\boldsymbol{x}_i^{(T)}-\boldsymbol{x}_l^{(T)}\|_2^2)\}
\label{eq:fit}
\end{align} 

After $T$ times heat diffusion, we assume there is enough information transfer. Thus, the updated signal of each node encodes sufficient topology information and can be used to improve the alignment between the two networks.

\subsection{Post-Processing}\label{sec:post}
In post-processing, we reassign the baseline alignment based on the signal similarity in $\X^{(T)}$. 
For each node $\nu_i \in \G_1$, we keep the $M$ most confident baseline alignments that connect $\nu_i$ to $M$ candidate nodes in $\G_2$. Thus, we only compare $\boldsymbol{x}_i^{(T)}$ with $M$ diffused signals and take the closest node as the updated alignment. By doing so, we can both increase the alignment accuracy and maintain a short alignment time.

\subsection{Complexity Analysis}\label{sec:comp}
    
\textbf{Baseline Alignment}: The computation complexity of the baseline alignment varies depending on the network embedding method or other network alignment approaches used as preliminary results to feed into our framework. 
    
\textbf{Alignment Sampling}: To sample the baseline alignment, we take $\gamma \leq N$ confident alignment pairs, then we randomly split them into the anchor alignment set and training alignment set. It takes $\mathcal{O}(N)$ times to sample the pairs.
    
\textbf{Head Diffusion Process}: Eq.~\ref{eq:diffuse} takes $N\times N \times C$ times operation for the matrix multiplication, causing $\mathcal{O}(C N^2)$ time complexity. However, this simulation can be accelerated by parallel computing on GPU, and the computation complexity can be reduced to $\mathcal{O}(\max\{C,N\})$.
    
\textbf{Post-Processing}: 
We compare $M$ candidate node signals on $\G_2$ to $N$ node signals on $\G_1$. Since the diffused signal is represented by $C$ dimension vectors, this process takes $\mathcal{O}(MNC)$ complexity.

In conclusion, the total complexity of the heat-diffusion based graph alignment is $\mathcal{O}(T C N^2)$ based on a CPU implementation and $\mathcal{O}(M N C)$ based on a GPU implementation.

\section{Experiments}

\textbf{Experimental Setup.}  
All experiments are performed on an Intel(R) Xeon(R) CPU E5-2680v4 at 2.40GHz with 126GB RAM. We accelerate the heat diffusion process by CUDA 10.0 using PyTorch 1.1 (Python 3.7) on Tesla V100 (32GB). 

\textbf{Dataset.} We use a PPI network commonly used in the graph alignment community, a subgraph of human PPI network with 3890 proteins as nodes, and 76,584 edges to represent protein-protein interactions \cite{breitkreutz2007biogrid}. 

\textbf{Noise on graph.} 
We synthesize a network $\G_2$ from a known network $\G_1=\{\V_1,\E_1\}$. We permute the nodes and perturb the edges with noise level $p_s$. 
Assume $\G_1$ is associated with \emph{adjacency matrix} $\A_1$.
We take two steps to generate the \emph{adjacency matrix} $\A_2$ of $\G_2$ : we first multiply $\A_1$ by a random permutation matrix $\perm$: $\A_1^{perm}=\perm\A_1\perm^\top$;
we then remove edges in $\A_1^{perm}$ with probability $p_s$ without disconnecting any nodes and finally get $\A_2$.

\textbf{Evaluation Metrics.}
We adopt commonly used metrics for PPI NA: accuracy at top-1 and top-5 choices, edge correctness ($EC$), and symmetric substructure score ($S^3$). 

\textit{Accuracy} is calculated as the ratio of the number of correct alignments to the total number of groudtruth alignments. 
We consider both top-1 and top-5 accuracy (Acc@1 and Acc@5).
\textit{EC} is the percentage of edges in one network that are aligned to the edges in another \cite{kuchaiev2010topological}.
\textit{EC} can be viewed as the measure of how much the edges are conserved between the  two graphs. 
$S^3$ is proposed in \cite{saraph2014magna} with respect to both the source network and the target network. It penalizes the alignment mapping of denser regions to sparse regions and vise-versa.

\textbf{Learnable heat diffusion on baseline alignments.}
To assess the performance of our method EDNA, we feed two different features, ndegree feature and struc2vec embedding, of PPI networks into our pipeline and calculate baseline alignment on them. struc2vec~\cite{ribeiro2017struc2vec} takes a hierarchy to measure node similarity at different scales for node embedding and preserves \emph{structural} similarity of nodes. ndegree features are calculated using the degree information of every node's neighbors considering neighbors up to k hops from the original node \cite{heimann2018regal}.

Also, the alignment from REGAL, a state-of-the-art NA method, is directly used as a baseline to show the capability of our method as a gerneralizable framework to boost performance of currently available alignments.

As shown in Tab. \ref{tab:acc}, EDNA drastically increases the performance (especially Acc@1) of baselines from different embedding methods. Taking Acc@1 as an example, EDNA reaches $92.6\%$ from $65.0\%$ for ndegree. Remarkably, our proposed method can significantly improve the performance of NA method REGAL from $59.6\%$ to $92.6\%$.

\begin{table}[htb]
\label{acc}
    \centering
    \small
  \begin{tabular}[b]{|l|cccc|}
    \hline
{\textbf{Method}}&{Acc@1}&{Acc@5}&{\textit{EC}}&{$S^3$}\\ \hline
\multicolumn{5}{|c|}{Baseline}\\
      \hline
     Struc2vec  & 63.9 & 88.7 & 78.6&75.7  \\
     REGAL  & 59.6 & 77.4& 49.3   &46.3    \\ 
     ndegree & 65.0 & 80.9 & 63.1  & 60.3\\
      \hline
      
\multicolumn{5}{|c|}{After EDNA} \\
      \hline
     Struc2vec  & 73.4& 95.4 & 55.4 &55.6  \\
     REGAL  & 92.6 &96.5   & 97.7   &98.5    \\ 
     ndegree & 92.6 & 96.7 & 98.1  & 98.8\\
     
    \hline
    \end{tabular}
 \vspace{-10pt}
    \caption{\textbf{Accuracy of baseline alignments and EDNA for fixed $5\%$ noise level. }}
    \vspace{-10pt}
    \label{tab:acc}
\end{table}

\begin{figure}[h]
    \centering
    \includegraphics[width=7.5cm]{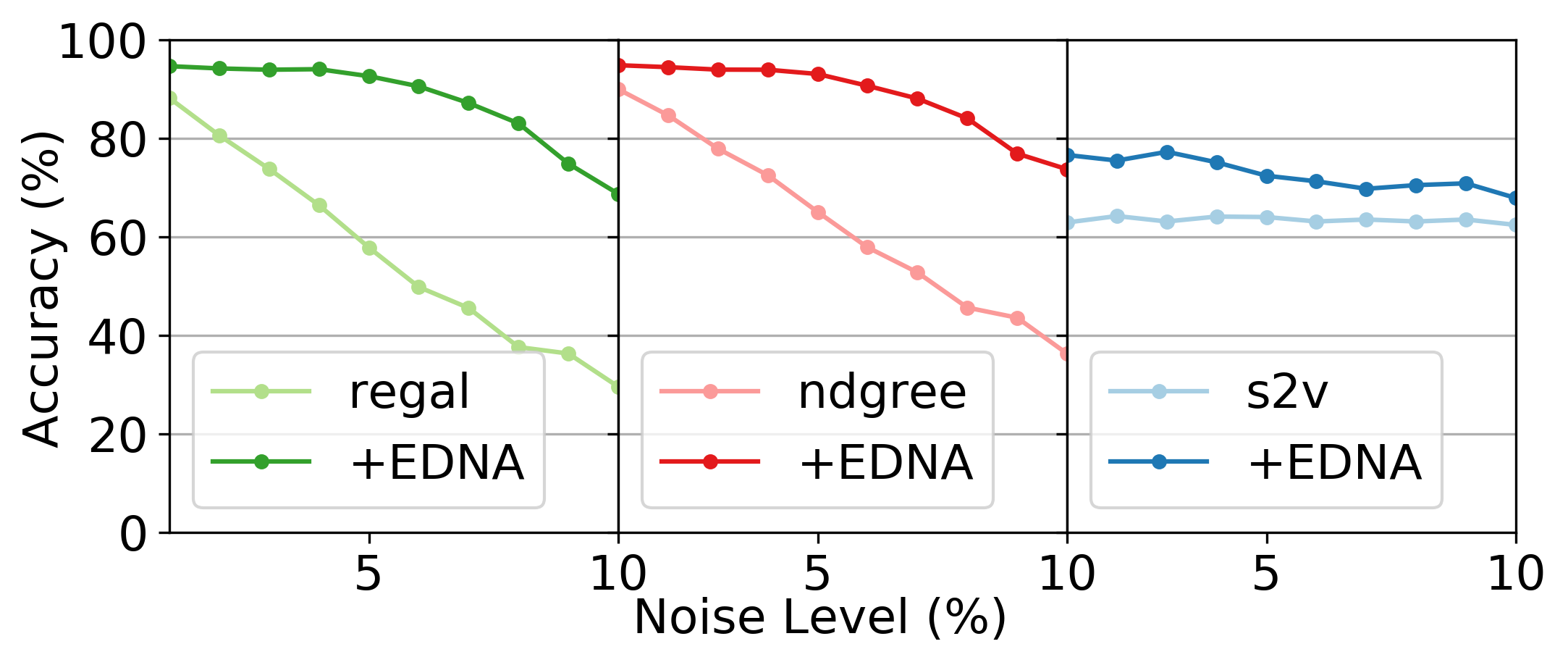}
    \small
     \vspace{-10pt}
    \caption{\textbf{Accuracy of baseline alignments and EDNA with varying noise levels from 1\% to 10\%}. The implementation with EDNA consistently outperforms baselines using different features from (struc2vec and ndegree) and NA method (REGAL).} 
    \label{fig:noise}
\end{figure}

Then, we investigate the performance of baselines compared to that after deploying our EDNA framework in terms of accuracy under different noise levels from 1\% up to 10\%. Fig. \ref{fig:noise} clearly shows that EDNA is consistently more accurate than baselines from different representations and NA methods with improved robustness, a testimony to its great potential for boosting NA performance.


\textbf{Scalability with Dataset Size.}
On Erdös-Rényi networks with n= 500 to 8000 nodes and a constant average degree of 10, we test the scalability of EDNA. Notably, the diffusion process can be run on GPU's in parallel, significantly reducing the time cost. In Tab. \ref{tab:scale}, we show that the running time on CPU or GPU increases sub-quadratically with the size of the networks. 

\begin{table}[htb]
    \centering
    \small
  \begin{tabular}[b]{l cccccccc }
    \toprule
  \textbf{Node Number}   &500& 1000 & 2000& 4000& 8000 \\
     \midrule
      {Top 1 Acc.} &67.0& 65.4 & 62.0& 58.2& 56.2 \\
      {Top 5 Acc.} &76.0& 70.8 & 67.5& 62.5& 58.4 \\

    \hline
      {CPU (mean)}   &17.8& 52.5 & 193& 795& 3041 \\
      
      {CPU (std)}   &1.53& 10.5 & 12.4& 62.1& 18.3 \\

    \hline
      {GPU (mean)}  &4.85& 10.9 & 34.5& 135& 272 \\
      
      {GPU (std)}  &0.46& 0.17 & 0.59& 3.7& 1.4 \\
      
     \bottomrule
    \end{tabular}
 \vspace{-10pt}
    \caption{\textbf{Scalability of EDNA with network size.} The alignment accuracy (\%) and running time (ms) is calculated for CPU or GPU respectively, using synthetic network with increasing number of nodes. With increasing network size the alignment accuracy decreases slightly, while the running time on CPU or on GPU increases. However, the simulation of the heat diffusion process on a GPU is much faster.}
         \vspace{-10pt}
    \label{tab:scale}
\end{table}

\textbf{Ablation Study.}
We ablate the main components of EDNA on PPI networks with noise level $10\%$, including anchor-based diffusion and evolutionary algorithm. Tab. \ref{ablation} reports the results with each component disabled for five metrics. With only the diffusion block, Acc@1 can reach $52.75\%$ from $35.63\%$, while using evolutionary diffusion, we can increase performance to $77.89\%$. Those two blocks are essential for maximizing alignment performance.
\begin{table}[!h]
    \centering
    \small
  \begin{tabular}[b]{lcccc}
    \toprule
{\textbf{Method}}&{Acc@1}&{Acc@5}&{\textit{EC}}&{$S^3$}\\ \midrule
     ndegree  & 35.63 & 57.09 & 27.77  &23.53  \\
     Diffusion  & 52.75 & 73.54 & 56.96 & 56.26    \\ 
     EDNA & 77.89 & 86.58 & 70.22 & 71.29\\
      \bottomrule
    \end{tabular}
 \vspace{-10pt}
    \caption{\textbf{Ablation study of EDNA.} We use ndegree as a baseline alignment method to compare with the performance after applying diffusion and full EDNA. We report the results of four metrics respectively. $p_s$ is set to $10\%$.}
     \vspace{-20pt}
      \label{ablation}
\end{table}



\section{Conclusion}
In this paper, we present a learnable evolutionary heat diffusion based algorithm for network alignment. On a general note, techniques for network alignment have a broad range of applications in computational biology. For example, in our hands, EDNA can be advantageous for integrating single cell RNA sequencing data, since single cells can be represented as nodes in a high-dimensional expression space, connected using k-nearest neighbor graphs. 


\nocite{langley00}

\bibliographystyle{icml2020}
\bibliography{example_paper}

\appendix

\end{document}